\documentclass[conference,10pt]{IEEEtran}
\IEEEoverridecommandlockouts
\usepackage{cite}
\usepackage{amsmath,amssymb,amsfonts}
\usepackage{algorithmic}
\usepackage{graphicx}
\usepackage{textcomp}
\usepackage{xcolor}
\usepackage{moresize}
\usepackage{multirow,threeparttable,booktabs,adjustbox}
\usepackage[ruled,vlined]{algorithm2e}
\usepackage[
    colorlinks,
    linkcolor=red,
    anchorcolor=blue,
    citecolor=green,
    bookmarks=false,
    ]{hyperref}

\graphicspath{{imgs/}}

\def\BibTeX{{\rm B\kern-.05em{\sc i\kern-.025em b}\kern-.08em T\kern-.1667em\lower.7ex\hbox{E}\kern-.125emX}}

\begin{document}

\title{CoopASD: Cooperative Machine Anomalous Sound Detection with Privacy Concerns}

\author{\IEEEauthorblockN{
Anbai Jiang\IEEEauthorrefmark{2}, 
Yuchen Shi\IEEEauthorrefmark{2}, 
Pingyi Fan\IEEEauthorrefmark{1}\IEEEauthorrefmark{2},
Wei-Qiang Zhang\IEEEauthorrefmark{2} and Jia Liu\IEEEauthorrefmark{2}\IEEEauthorrefmark{3}}\\

\vspace*{-4mm}
\IEEEauthorblockA{
\IEEEauthorrefmark{1}Email: fpy@tsinghua.edu.cn\\
\IEEEauthorrefmark{2}Department of Electronic Engineering, BNRist, Tsinghua University, Beijing, China.\\
\IEEEauthorrefmark{3}Huakong AI Plus, Beijing, China.}}

\maketitle

\begin{abstract}
Machine anomalous sound detection (ASD) has emerged as one of the most promising applications in the Industrial Internet of Things (IIoT) due to its unprecedented efficacy in mitigating risks of malfunctions and promoting production efficiency. Previous works mainly investigated the machine ASD task under centralized settings. However, developing the ASD system under decentralized settings is crucial in practice, since the machine data are dispersed in various factories and the data should not be explicitly shared due to privacy concerns. To enable these factories to cooperatively develop a scalable ASD model while preserving their privacy, we propose a novel framework named CoopASD, where each factory trains an ASD model on its local dataset, and a central server aggregates these local models periodically. We employ a pre-trained model as the backbone of the ASD model to improve its robustness and develop specialized techniques to stabilize the model under a completely non-iid and domain shift setting. Compared with previous state-of-the-art (SOTA) models trained in centralized settings, CoopASD showcases competitive results with negligible degradation of 0.08\%. We also conduct extensive ablation studies to demonstrate the effectiveness of CoopASD.
\end{abstract}


\section{Introduction}

The Industrial Internet of Things (IIoT) is a specialized form of the Internet of Things (IoT) for industrial applications, which has emerged as a driving force for the evolution of conventional manufacturing into a digital era. The key idea of IIoT is to harness the power of big data for advanced automation and optimization, by seamlessly integrating data collection, data transmission, and data analysis in an automated pipeline, where the infrastructure of IoT paves the way for reliable data collection and transmission, and machine learning (ML) algorithms enable it to extract valuable insights from massive volumes of production data, bringing forth unprecedented opportunities for downstream applications.

Machine anomalous sound detection (ASD) is one of the most emerging tasks in IIoT, which seeks to detect machine malfunctions when only audio of normal working status is provided. Compared with fault detection tasks, the ASD task emphasizes the absence of labeled anomalies for training, which is more applicable in real production sites, since recorded malfunctions are scarce and are commonly used as validation for detection models, while data of normal working status can be readily collected. A well-performing machine ASD model is sure to significantly enhance operational efficiency, minimize downtime, and mitigate risks associated with machine failures and malfunctions.

The machine ASD task has been widely studied under centralized settings in recent years~\cite{han2024exploring,wilkinghoff2024self,zhang2024dual,hou2023decoupling,jiang2023unsupervised,liu2022anomalous,wilkinghoff2021sub}, where machine data are first aggregated on a central server before training the ASD model. However, the centralized paradigm may not be directly applicable in real scenarios, especially for small-scale factories with a limited number of machines. In common scenarios, each factory keeps a local dataset consisting of normal machine audio and minor anomalous audio, where the anomalies are only used for validation. For each of these factories, neither the quality of the training data nor the number of labeled anomalies is sufficient to develop a scalable ASD model, and the ASD model will be easily overfitted if it is built only on the local dataset. On the other hand, if these factories opt to cooperate with each other and leverage all available data, they are still capable of developing a well-performing and scalable ASD system, since both the diversity of the training data and the robustness of the validation data are improved through cooperation.

Nevertheless, training a unified ASD model in decentralized settings incurs two critical issues:

\begin{enumerate}
    \item Non-iid data. The machine types will likely vary across different factories, although the intrinsic patterns of all possible malfunctions may be similar in semantics.
    \item Data privacy. Business secrets such as parameter settings and production schedules can be readily inferred from the machine data. Thus the machine data should not be shared across factories.
\end{enumerate}

To tackle these problems, we propose CoopASD, a novel framework that seeks to develop a unified and well-performing machine ASD model for dozens of small factories via cooperation while preserving privacy. CoopASD follows the architecture of FedAvg~\cite{mcmahan2017communication}, where factories are considered as local clients and a central server aggregates all local updates. In the proposed scheme, each factory trains a local ASD model on its own dataset and periodically uploads the local model to the central server, while the central server aggregates these local models, updates the global model and broadcasts the updated global model to factories. The local training process is similar to the training processes in centralized settings~\cite{han2024exploring}. To alleviate the convergence problem induced by the non-iid data and the overfitting problem induced by the absence of labeled anomalies, three regularization methods are adopted in the local training process, namely sampling, selective upload and early stop. It is noted that machine data are not transferred between factories in both the training and inference stages, thus preventing privacy leakage.

The experiment is conducted on the dataset of DCASE 2023 Task 2~\cite{Dohi_arXiv2023_01} in a completely non-iid and domain shift setting, where each factory has a unique machine type. CoopASD demonstrates competitive results on all 14 machine types, with minor degradation of 0.08\% compared with the state-of-the-art (SOTA) models in centralized settings~\cite{JieIESEFPT2023,LvHUAKONG2023,han2024exploring,wilkinghoff2024self,zhang2024dual}. We also conduct extensive ablation studies to showcase the efficacy of the modifications. To the best of our knowledge, we are the first to explore the machine ASD task under decentralized settings.

The main contributions of CoopASD can be summarized:

\begin{enumerate}
    \item We propose CoopASD, a novel framework that enables factories to cooperatively develop a unified ASD model when no anomalies are presented for training.
    \item CoopASD combines the machine data and computation resources of all factories while preserving privacy.
    \item Regularization methods are adopted to stabilize CoopASD, enabling it to converge in a completely non-iid and domain shift setting.
    \item The performance of CoopASD is comparable with SOTA models under centralized settings with minor degradation of 0.08\%.
\end{enumerate}

\section{Related Work}

\subsection{Machine ASD in Centralized Settings}

A typical ASD model can be decomposed into a feature extractor and an anomaly detector, where the feature extractor extracts semantic representations of the machine audio, and the anomaly detector processes the audio representation and outputs a high anomaly score if it is anomalous. Models can be roughly divided into feature-centric models and anomaly-centric models depending on the emphasis.

\textbf{Feature-centric models} aim to extract semantic-rich and robust representations for machine audio, while adopting shallow anomaly detectors, such as Gaussian mixture model (GMM) and k-nearest neighbor (KNN)~\cite{ramaswamy2000efficient}. Liu et~al.~\cite{liu2022anomalous} proposed STgram which trains a dual path network to extract features from two perspectives. Zhang et~al.~\cite{zhang2024dual} extended the STgram by employing three sub-networks to extract features from four hierarchies. Wilkinghoff et~al.~\cite{wilkinghoff2024self} utilized the inner consistency of two sub-networks to derive robust representations. Han et~al.~\cite{han2024exploring} explored the usage of large pre-trained models for ASD.

\textbf{Anomaly-centric models} explore novel ML-based anomaly detectors where the audio representations are often spectrograms. Autoencoder~\cite{Dohi_arXiv2023_01} detects anomalies by the reconstruction error of spectrograms, where anomalies are expected to have bigger reconstruction error after training the network on normal spectrograms. Jiang et~al.~\cite{jiang2023unsupervised} addressed the denoising problem of autoencoder by introducing a discriminator to provide more reliable gradients. Besides the autoencoder, flow model~\cite{dohi2021flow} learns the distribution of normal audio representations and predicts the likelihood of each audio clip.


\subsection{Anomaly Detection in Decentralized Settings}

Multiple anomaly detection approaches have been proposed under the framework of federated learning (FL). Li et~al.~\cite{li2019abnormal} detected malicious clients of a FL system by reconstructing the local model weights updates. Nguyen et~al.~\cite{nguyen2019diot} proposed D{\"I}oT which monitors the packet transmission of an IoT network and detects malicious devices, by modeling the likelihood of the packet sequence.

\section{Proposed Method}
\label{sec:method}

\begin{figure}[t]
    \centering
    \includegraphics[width=0.95\linewidth]{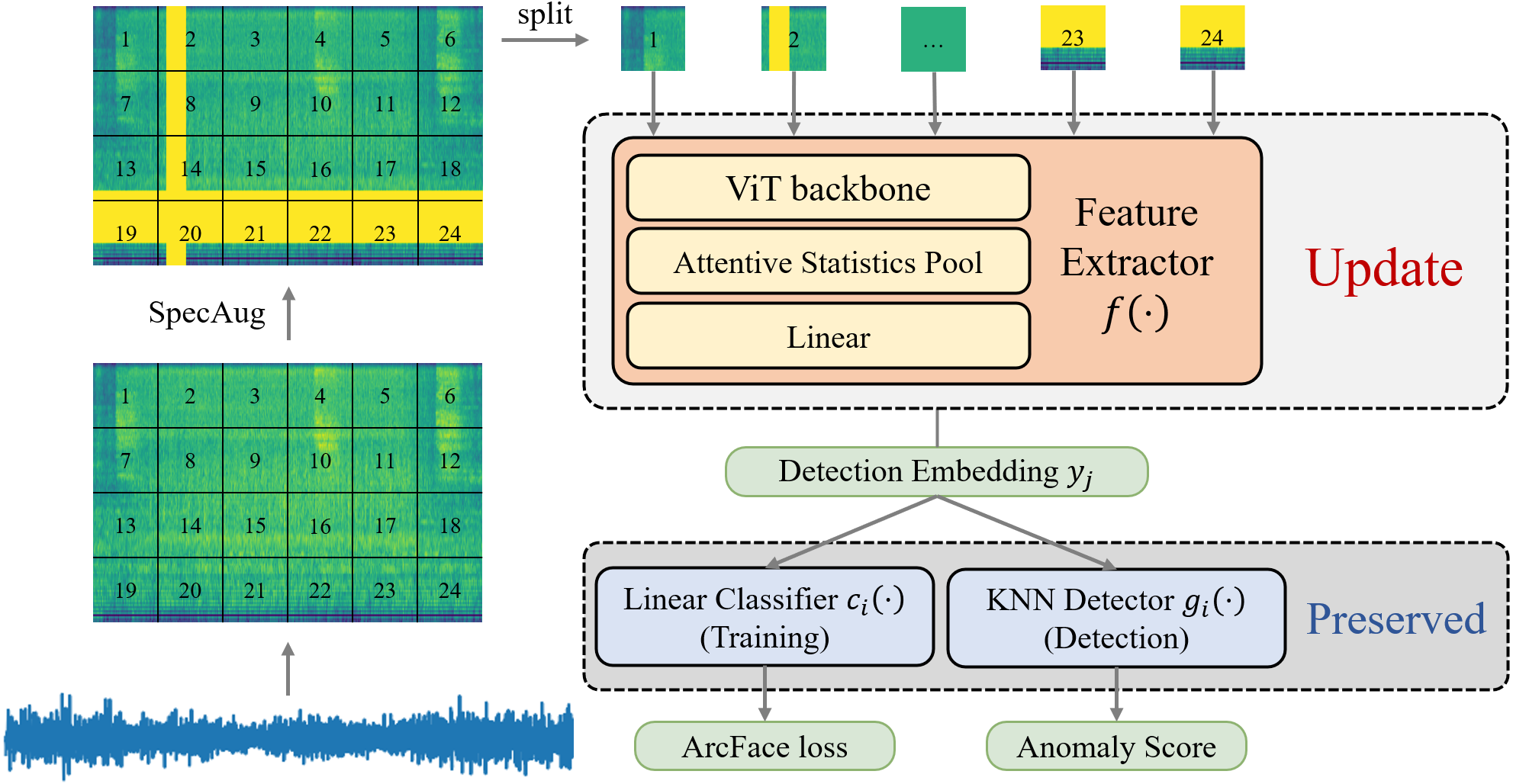}
    \caption{Architecture of the ASD Model in CoopASD. The feature extractor $f(\cdot)$ is updated globally and shared among factories, while the linear classifier $c_i(\cdot)$ and KNN detector $g_i(\cdot)$ are uniquely constructed and preserved locally.}
    \label{fig:model}
\end{figure}

This section introduces CoopASD in a bottom-up order.

\subsection{ASD Model}
\label{subsec:asd_model}

The ASD model of factory $i$ consists of a feature extractor $f(\cdot)$ and an anomaly detector $g_i(\cdot)$, where $f(\cdot)$ is shared across factories and $g_i(\cdot)$ is constructed locally. Fig.~\ref{fig:model} presents the general structure of the ASD model. For each normal recording $x_j$ from the local dataset of factory $i$, it is first converted to a log-mel spectrogram, then sent to the feature extractor $f(\cdot)$. SpecAug~\cite{park19e_interspeech} is applied to the spectrogram which masks a portion of the spectrogram to improve the robustness. The feature extractor $f(\cdot)$ adopts a ViT~\cite{dosovitskiy2021an} backbone, which splits the spectrogram into patches, encodes each patch as an embedding by a linear layer, and processes them by stacks of Transformer~\cite{vaswani2017attention} blocks, outputting a series of patch features. An attentive statistical pooling layer~\cite{dawalatabad21_interspeech} is appended to the ViT backbone to fuse these patch features into an utterance embedding $u_j$, and a linear layer is employed to map $u_j$ to a low-dimensional detection embedding $y_j$, which is further processed by the anomaly detector $g_i(\cdot)$. To improve the robustness, the ViT backbone is initialized from BEATs~\cite{pmlr-v202-chen23ag}, a pre-trained ViT model for audio classification.

The anomaly detector $g_i(\cdot)$ of factory $i$ is a simple KNN detector. A local memory bank $M_i$ of factory $i$ is first set up by the embeddings of the local training dataset $\mathcal{D}_i^{train}$:

\begin{equation}
    M_i=\left\{y_j=f(x_j) \mid x_j{\in}\mathcal{D}_i^{train}\right\}
\label{eq:memory}
\end{equation}
Since $\mathcal{D}_i^{train}$ only consists of normal audio, $M_i$ serves as a set of normality templates in the feature space. For each query embedding $y_q=f(x_q)$ of the local test dataset $\mathcal{D}_i^{test}$, $g_i(\cdot)$ infers a subset $N_{i,q}^{(k)}$ of $M_i$, which consists of the top-$k$ closest embeddings of $M_i$ to $y_q$:

\begin{equation}
    N_{i,q}^{(k)}=\mathop{\arg\min}_{N{\subset}M_i,\left|N\right|=k}\sum\limits_{y_j{\in}N}\frac{y_j^Ty_q}{{\lVert y_j \rVert}_2{\cdot}{\lVert y_q \rVert}_2}
\end{equation}
where cosine distance is adopted as the distance metric. The anomaly score is defined as the mean distance of $N_{i,q}^{(k)}$ to $y_q$:

\begin{equation}
    g_i(y_q)=\frac{1}{k}\sum\limits_{y_j{\in}N_{i,q}^{(k)}}\left(1-\frac{y_j^Ty_q}{{\lVert y_j \rVert}_2{\cdot}{\lVert y_q \rVert}_2}\right)
\label{eq:ano_score}
\end{equation}

\subsection{Local Training Process}
\label{subsec:local_training}

\begin{figure}[t]
    \centering
    \includegraphics[width=0.7\linewidth]{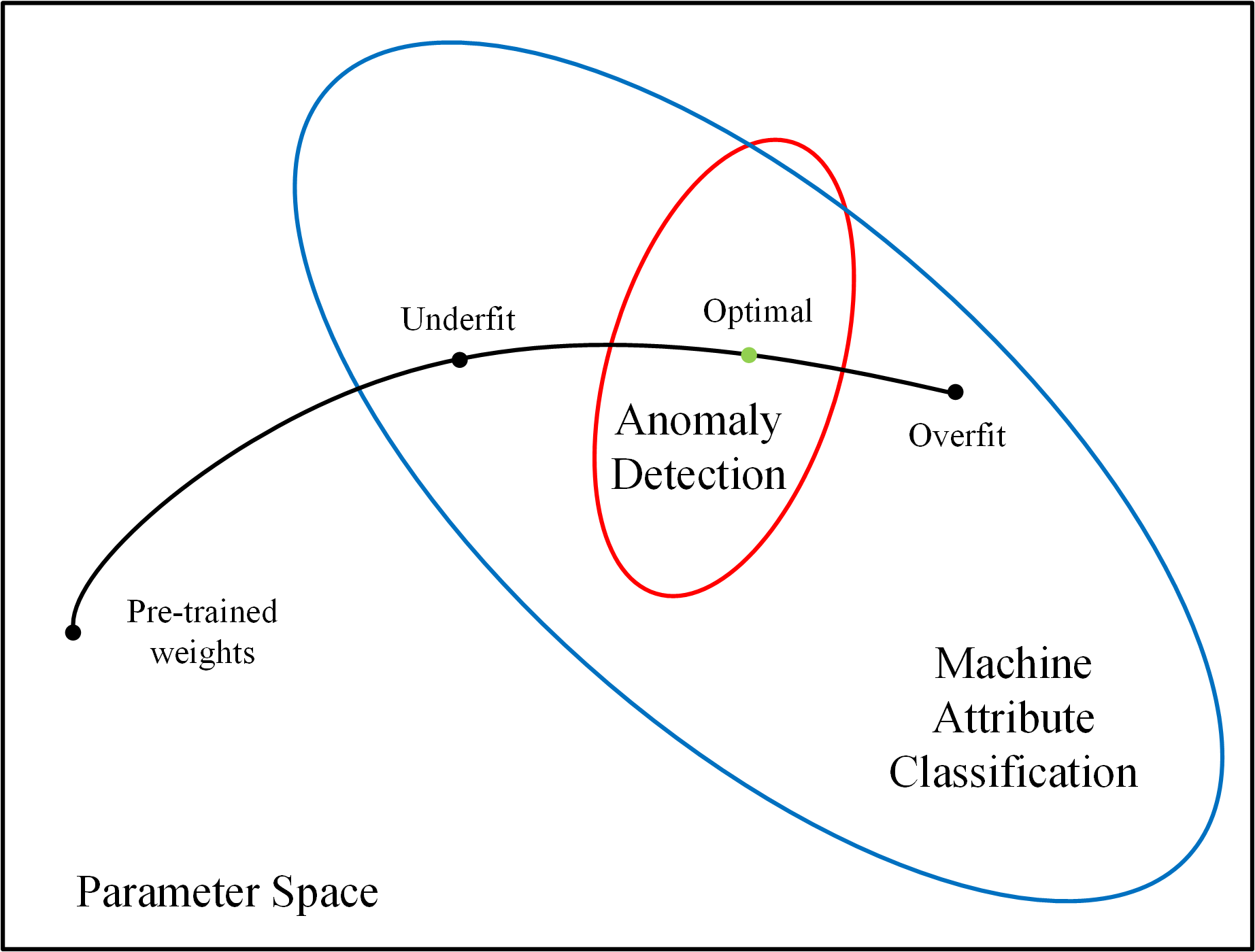}
    \caption{Training process viewed in parameter space. The red and blue ovals denote the sweet spots for anomaly detection and machine attribute classification respectively. The ASD model is trained by classifying machine attributes since labeled anomalies are not provided for training. Therefore, one can not tell when to stop training, and the model is likely to be overfitted or underfitted.}
    \label{fig:ft}
\end{figure}

Each factory trains an ASD model on its own dataset and periodically uploads the local model to the central server. Since labeled anomalies are not provided for training, the ASD model is trained by classifying the attributes of machine working conditions, such as speed, operation voltage and rotation velocity. These attributes are handy for collection, and each unique combination of attributes is considered a new label. A simple linear classifier $c_i(\cdot)$ is appended to the feature extractor $f(\cdot)$ for each factory $i$, which maps the output of $f(\cdot)$ to the local class labels. Since attributes of different factories are completely different, the linear classifier $c_i(\cdot)$ only predicts all locally available labels of $\mathcal{D}_i^{train}$ and is not uploaded to the central server.

Since the number of available attributes is always limited for each factory, the model can easily predict these attributes after quick adaptation. To further enforce the classification task, ArcFace loss~\cite{deng2019arcface} is adopted in CoopASD instead of cross-entropy loss, which further restricts the decision zones:

\begin{equation}
    L=-\frac{1}{N}\sum\limits_{j=1}^N\log\frac{e^{s\cos(\theta_{l_j}+m)}}{e^{s\cos(\theta_{l_j}+m)}+\sum\limits_{c=1,c{\neq}l_j}^{C_i}e^{s\cos\theta_{c}}}
\label{eq:loss}
\end{equation}
where $l_j$ is the label of $y_j$, $C_i$ is the number of classes of $\mathcal{D}_i^{test}$, and $s$ and $m$ are two hyperparameters that constrain the decision zones. $\theta_c$ is the angle between $y_j$ and the registered embedding of the $c$-th class, which is the $c$-th column of the weight $W$ of the linear classifier $c_i(\cdot)$:

\begin{equation}
    \theta_c=\arccos\left(\frac{W_c^Ty_j}{{\lVert W_c \rVert}_2{\cdot}{\lVert y_j \rVert}_2}\right)
\end{equation}

\begin{figure}[t]
    \centering
    \includegraphics[width=0.95\linewidth]{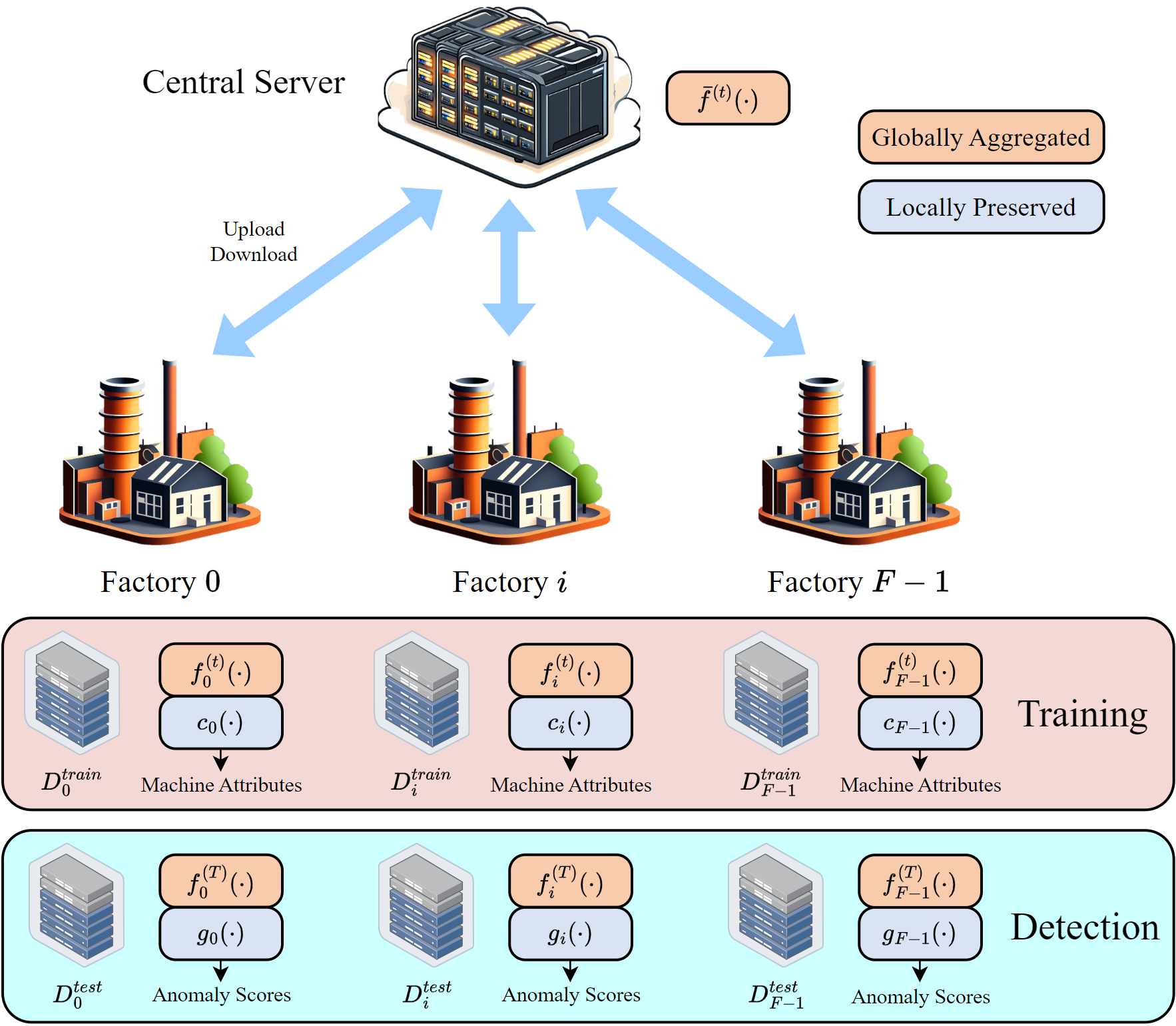}
    \caption{Training and detection procedures of CoopASD.}
    \label{fig:framework}
\end{figure}

\subsection{Global Aggregation Process}

Following the framework of FedAvg~\cite{mcmahan2017communication}, CoopASD trains the ASD model in a decentralized setting, by alternating between the local training processes of factories and the global aggregation process of the central server. However, directly applying vanilla FedAvg for the machine ASD task yields unsatisfactory results, which is due to the convergence problem induced by the completely non-iid data and the overfitting problem induced by the absence of labeled anomalies for training. As introduced in Section~\ref{subsec:local_training}, the ASD model is trained by classifying machine attributes. On the one hand, since each factory has a unique machine type, the problem is completely non-iid, where not only the machine audio but also the attributes are completely different for each factory. This incurs a severe convergence problem for the ASD model. On the other hand, as depicted in Fig.~\ref{fig:ft}, the classification accuracy of machine attributes is not a valid indicator of ASD performance, and the ASD model can be easily overfitted (in most cases) or underfitted. This calls for a delicate scheduler for the alternation between the local training processes and the global aggregation process.

To tackle these problems, three modifications are made in the proposed scheme, namely sampling, selective upload and early stop. Firstly, since applying sampling strategies to clients has been proven effective in recent literature~\cite{shi2024sam}, a uniform sampler is adopted to select factories participating in the current global epoch with a probability of $p$. Secondly, since the data are completely non-iid, the local linear classifiers of different factories yield distinct decision zones after local training. If a unified classifier is adopted for all factories, the model has to be updated frequently to ensure convergence, which imposes huge burdens on the communication network. Therefore, only the feature extractor $f(\cdot)$ is uploaded and aggregated by the central server, while each linear classifier $c_i(\cdot)$ is maintained locally. Finally, to alleviate the overfitting problem, an early stop strategy based on classification accuracy is employed in the local training process, where a first-in-first-out queue $A_i$ records the classification accuracy of batches. If the mean of $A_i$ exceeds an accuracy threshold $\epsilon$, the local training process is early-stopped to prevent overfitting.

\begin{algorithm}[t] 
    \SetAlgoLined
    \KwIn{local training datasets $\left\{\mathcal{D}_i^{train}\right\}$, pre-trained ViT backbone $\theta_{v}$, global epoch $T$, local step $S$, sampling probability $p$ and accuracy threshold $\epsilon$}
    Initialize the ViT backbone of $\bar{f}_i^{(0)}$ by $\theta_{v}$\;
    \For{\textcolor{black}{$t=0,1,\cdots,T-1$}}{
        Uniformly sample a subset of factories $\mathcal{P}_{t}$ by $p$\;
        \For{each factory $i\in \mathcal{P}_t$ \textbf{in parallel}}{
            Update the feature extractor $f_i^{(t)}(\cdot){\leftarrow}\bar{f}^{(t)}(\cdot)$\;
            Initialize the accuracy queue $A_i$\;
            \For{$s=0,1,\cdots,S-1$}{
                Sample a mini-batch $\xi_s^{(i)}{\subset}D_i^{train}$\;
                Train $f_i^{(t)}(\cdot)$ and $c_i(\cdot)$ on  $\xi_s^{(i)}$ by Eq.~\eqref{eq:loss}\;
                Update $A_i$\;
                \uIf{$\text{mean}(A_i)>\epsilon$}{
                    break\; 
                }
            }
            Upload $f_i^{(t)}(\cdot)$ to the central server.
        }
        Server updates the feature extractor by Eq.~\eqref{eq:glob_up}\;
    }
    \KwOut{global feature extractor $\bar{f}^{(T)}(\cdot)$}
    \caption{Training}
    \label{algo:train}
\end{algorithm}

\begin{algorithm}[t] 
    \SetAlgoLined
    \KwIn{local training datasets $\left\{\mathcal{D}_i^{train}\right\}$, local test dataset $\left\{\mathcal{D}_i^{test}\right\}$, number of neighbors $k$}
    \For{each factory $i=0,1,\cdots,F-1$ \textbf{in parallel}}{
        Fetch the global feature extractor $\bar{f}^{(T)}(\cdot)$\;
        Construct local memory bank $M_i$ by Eq.~\eqref{eq:memory}\;
        \For{each query $x_q\in \mathcal{D}_i^{test}$}{
            Select the set of top-k closest neighbors $N_{i,q}^{(k)}$\;
            Infer anomaly score $g_i\left(\bar{f}^{(T)}(x_q)\right)$ by Eq.~\eqref{eq:ano_score}\;
        }
    }
    \KwOut{anomaly scores $\left\{g_i\left(\bar{f}^{(T)}(x_q)\right) \mid x_q{\in}\mathcal{D}_i^{test},\ i=0,1,\cdots,F-1\right\}$}
    \caption{Detection}
    \label{algo:test}
\end{algorithm}

Fig.~\ref{fig:framework} illustrates the general framework of CoopASD, and the training procedure is depicted in Algorithm~\ref{algo:train}. For global epoch $t$, we first uniformly sample a subset of factories $\mathcal{P}_{t}$ with probability $p$, and factories not in $\mathcal{P}_{t}$ will not participate in the current epoch. For each participating factory $i{\in}\mathcal{P}_{t}$, it first fetches the newest global feature extractor $\bar{f}^{(t)}(\cdot)$ as the new local feature extractor $f_i^{(t)}(\cdot)$, then trains $f_i^{(t)}(\cdot)$ and $c_i(\cdot)$ on $\mathcal{D}_i^{train}$ by Equation~\ref{eq:loss}. The local model is trained for $S$ steps at most, and if the mean of the accuracy queue $A_i$ exceeds the threshold $\epsilon$, the local training process is early-stopped. When all participants have finished the local training process, the central server aggregates the local feature extractors of all participants $\left\{f_i^{(t)}(\cdot) \mid i{\in}\mathcal{P}_{t}\right\}$, and updates the global feature extractor by the average of the weights of these local models. Let $\bar{\theta}^{(t+1)}$ and $\theta_i^{(t)}$ denote the weights of $\bar{f}^{(t+1)}(\cdot)$ and $f_i^{(t)}(\cdot)$ respectively. The global aggregation process is formulated as:

\begin{equation}
    \bar{\theta}^{(t+1)}=\frac{1}{\left|\mathcal{P}_t\right|}\sum\limits_{i{\in}\mathcal{P}_t}\theta_i^{(t)}
\label{eq:glob_up}
\end{equation}
where $\left|\mathcal{P}_t\right|$ is the number of participating factories.

Algorithm~\ref{algo:test} presents the anomaly detection procedure of $F$ factories. For factory $i$, it downloads the final global feature extractor $\bar{f}^{(T)}(\cdot)$ and conducts anomaly detection locally. It is noted that both the training and detection procedures do not share any machine data and audio features across factories, thus the privacy of these factories can be preserved.

\begingroup
\setlength{\tabcolsep}{2pt}
\renewcommand{\arraystretch}{1.3}
\begin{table*}[th]
  \caption{ASD Performance on the DCASE 2023 dataset}
  \label{tab:dcase23_compar}
  \centering
  \scriptsize
  \adjustbox{max width=\textwidth}{
  \begin{tabular}{ccccccccc|c|ccccccc|c|c}
    \toprule
    \multirow{2}*{Settings} & \multirow{2}*{Models} & \multicolumn{8}{c}{Development set} & \multicolumn{8}{c|}{Evaluation set} & All \\
    & & bearing & fan & gearbox & slider & ToyCar & ToyTrain & valve & hmean & bandsaw & grinder & shaker & ToyDrone & ToyNscale & ToyTank & Vacuum & hmean & hmean \\
    \midrule
    \multirow{5}*{Centralized} & No. 1~\cite{JieIESEFPT2023} & 64.41 & \textbf{76.27} & \textbf{74.78} & \textbf{91.83} & 51.66 & 53.17 & 85.44 & \textbf{68.11} & 60.97 & 65.18 & 63.50 & 55.71 & \textbf{84.92} & 60.72 & 92.27 & 66.97 & 67.54 \\
    & No. 2~\cite{LvHUAKONG2023} & \textbf{72.39} & 62.41 & 74.41 & 87.84 & 59.10 & 58.67 & 65.53 & 67.38 & 55.47 & 64.76 & 70.98 & 52.89 & 71.90 & 70.73 & 91.48 & 66.39 & 66.88 \\
    & Han et~al.~\cite{han2024exploring} & 57.10 & 62.76 & 67.52 & 79.11 & \textbf{63.47} & 57.35 & \textbf{67.79} & 64.31 & - & - & - & - & - & - & - & - & -\\
    & FeatEx~\cite{wilkinghoff2024self} & - & - & - & - & - & - & - & 66.95 & - & - & - & - & - & - & - & 68.52 & \textbf{67.73} \\
    & Zhang et~al.~\cite{zhang2024dual} & - & - & - & - & - & - & - & - & - & - & - & - & - & - & - & 71.27 & - \\
    \midrule
    Decentralized & CoopASD (Ours) & 58.29 & 62.07 & 67.08 & \textbf{97.77} & 62.04 & \textbf{61.48} & 50.39 & 63.28 & \textbf{65.75} & \textbf{68.60} & \textbf{78.12} & \textbf{57.29} & 84.40 & \textbf{71.07} & \textbf{96.36} & \textbf{72.66} & 67.65 \\
    \bottomrule
  \end{tabular}}
\end{table*}
\endgroup

\addtolength{\topmargin}{0.01in}  

\section{Experiment}

\subsection{Dataset}

The experiment is conducted on the dataset of DCASE 2023 Task 2, a dedicated machine ASD dataset with 14 machine types. The dataset can be divided into a development set and an evaluation set, each of which consists of 7 machine types. For each machine type, the training set consists of 1000 normal clips, while the test set contains 100 normal clips and 100 anomalous clips. Each training clip is accompanied by multiple attributes of the working conditions, which are utilized as the labels of the deputy task. The experiment is conducted in a completely non-iid and domain shift setting, where each factory corresponds to a unique machine type, resulting in 14 factories. The performance is measured by the area under the receiver operating characteristic (ROC) curve (AUC) and the partial-AUC (pAUC) following the challenge rules~\cite{Dohi_arXiv2023_01}. We report the harmonic mean of all AUC and pAUC for each machine type, and a harmonic mean of the whole dataset.

It is noted that the DCASE 2023 dataset also features domain shift, where the 1000 normal clips of each machine type can be further divided into 990 clips from a source domain and 10 clips from a target domain. However, the domain shift problem is not discussed in the proposed scheme. A soft scoring mechanism~\cite{jiang2023thuee} is employed based on the proposed KNN detector, where two KNN detectors are constructed based on the embeddings of source data and target data respectively, and the minimum of the scores given by two detectors is utilized as the final anomaly score.

\subsection{Hyperparameter Settings}

Following the settings of the pre-trained model~\cite{pmlr-v202-chen23ag}, we pad or truncate each clip to 10s, and calculate the log-mel spectrogram with 128 mel filters and a 25 ms window that shifts every 10 ms. For SpecAug, the maximum length of frequency masking and time masking is 80. The ASD model consists of 90M parameters. The ViT backbone consists of 12 Transformer blocks with 8 attention heads and a hidden size of 768, while the size of the detection embedding is 128. An AdamW~\cite{loshchilov2018decoupled} optimizer with a learning rate of 0.0001 is adopted for each factory and is reinitialized every global epoch. For each local training process, the local ASD model is trained by a batch of 32 samples with a gradient accumulation of 8 for 200 steps at most. The size of the accuracy queue $A_i$ is 10 and the threshold $\epsilon$ is 0.95. The sampling probability $p$ is 0.5 and the number of neighbors $k$ is 1.

\subsection{Experiment Results}

To the best of our knowledge, the machine ASD task under decentralized settings has not been explored in previous literature. Therefore, CoopASD is compared with five SOTA models under centralized settings, including the top two best-performing systems of the challenge~\cite{JieIESEFPT2023,LvHUAKONG2023}. All models are single models without ensembling, except for the second system of the challenge~\cite{LvHUAKONG2023}.

Table~\ref{tab:dcase23_compar} compares CoopASD with five SOTA models on the DCASE 2023 dataset. CoopASD achieves the best ASD results on 8 out of 14 machine types, with a harmonic mean of 63.28\% on the development set and 72.66\% on the evaluation set. The overall performance of CoopASD is 67.65\%, which is only 0.08\% lower than the best centralized model.

\begin{table}[t]
    \centering
    \caption{Comparison of the ASD model between centralized and decentralized settings}
    \label{tab:cen_compar}
    \begin{tabular}{ccccccc}
        \toprule
        \multirow{2}[3]{*}{Setting} & \multicolumn{3}{c}{Centralized} & \multicolumn{3}{c}{Decentralized} \\
        \cmidrule(lr){2-4} \cmidrule(lr){5-7}
        & dev & eval & all & dev & eval & all \\
        \midrule
        hmean & \textbf{64.24} & \textbf{74.23} & \textbf{68.87} & 63.28 & 72.66 & 67.65 \\
        \bottomrule
    \end{tabular}
\end{table}

\begin{table}[t]
    \centering
    \caption{Validation of the proposed techniques}
    \label{tab:tech_compar}
    \begin{tabular}{cccccc}
        \toprule
        Sampling & Selective upload & Early stop & dev & eval & all \\
        \midrule
        & & & 61.68 & 68.28 & 64.81 \\
        \checkmark & & & 61.38 & 71.66 & 66.12 \\
        \checkmark & \checkmark & & 61.76 & 71.67 & 66.35 \\
        \checkmark & \checkmark & \checkmark & \textbf{63.28} & \textbf{72.66} & \textbf{67.65} \\
        \bottomrule
    \end{tabular}
\end{table}

\begin{table}[!ht]
    \centering
    \caption{Comparison of cooperative training and solo training}
    \label{tab:coop_compar}
    \begin{tabular}{ccccccc}
        \toprule
        \multirow{2}[3]{*}{Setting} & \multicolumn{3}{c}{Cooperative} & \multicolumn{3}{c}{Solo} \\
        \cmidrule(lr){2-4} \cmidrule(lr){5-7}
        & dev & eval & all & dev & eval & all \\
        \midrule
        hmean & 63.28 & \textbf{72.66} & \textbf{67.65} & \textbf{65.02} & 69.96 & 67.40 \\
        \bottomrule
    \end{tabular}
\end{table}

\subsection{Ablation Study}

\subsubsection{Centralized versus Decentralized}

The competitive performance of CoopASD can be attributed to both the powerful ASD model and the well-designed iterative scheme. To showcase the efficacy of the two contributions respectively, CoopASD is compared with the same ASD model trained under a centralized setting, which is demonstrated in Table~\ref{tab:cen_compar}. Compared with Table~\ref{tab:dcase23_compar}, the centralized ASD model is 1.14\% better than previous ASD models, while switching to a decentralized setting only incurs degradation of 1.22\%.

\subsubsection{Regularization Methods}

Compared with FedAvg, three techniques are adopted in CoopASD to improve ASD performance, namely sampling, selective upload and early stop. Table~\ref{tab:tech_compar} validates the effectiveness of the proposed techniques, where the performance gradually increases from 64.81\% to 67.65\% when applying the proposed techniques progressively. This proves the efficacy of the techniques.

\subsubsection{Cooperative versus Solo}

Table~\ref{tab:coop_compar} compares the results of cooperative training and solo training, where solo training means each factory trains the model individually. Cooperative training yields a slightly better ASD model with an improvement of 0.15\%, which is probably due to the improved diversity of the training data and the underlying similarity of different malfunctions. More importantly, the ASD model obtained by cooperative training is much more robust, since it is unified and has been validated on multiple test sets. Therefore, factories are more willing to develop a unified ASD system through cooperation.

\subsection{Sampling Probability}

\begin{figure}[t]
    \centering
    \includegraphics[width=0.95\linewidth]{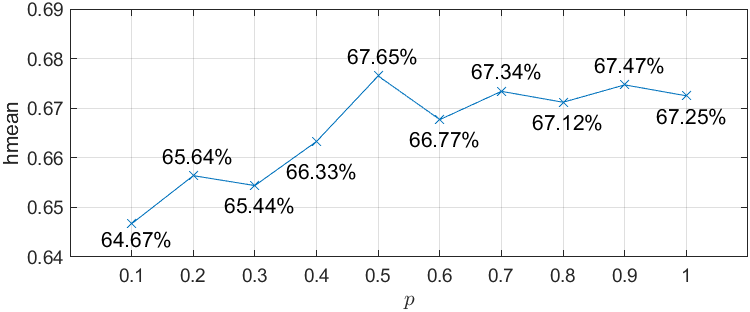}
    \caption{Comparison of different sampling probability}
    \label{fig:samp_prob_compar}
\end{figure}

Fig.~\ref{fig:samp_prob_compar} compares different sampling probability $p$, where bigger probability generally yields consistently better results, and the best result is achieved when $p=0.5$.




\section{Conclusion}

In this paper, we proposed CoopASD, a decentralized ASD framework that allows factories to cooperatively train a robust ASD model while preserving privacy. CoopASD is built on a powerful ASD model that leverages a pre-trained ViT backbone and is trained by alternating between the local training processes of factories and the global aggregation process of the central server. The experiment under a completely non-iid and domain shift setting demonstrated the effectiveness and the generalization capability of CoopASD. Our future work will focus on reducing the number of trainable parameters and analyzing the convergence of the ASD model theoretically.

\section*{Acknowledgment}

This work was supported by the National Key Research and Development Program of China (Grant NO.2021YFA1000500(4)) and the National Natural Science Foundation of China under Grant No. 62276153.

\bibliographystyle{IEEEtran}
\bibliography{refer.bib}

\begin{thebibliography}{10}
\providecommand{\url}[1]{#1}
\csname url@samestyle\endcsname
\providecommand{\newblock}{\relax}
\providecommand{\bibinfo}[2]{#2}
\providecommand{\BIBentrySTDinterwordspacing}{\spaceskip=0pt\relax}
\providecommand{\BIBentryALTinterwordstretchfactor}{4}
\providecommand{\BIBentryALTinterwordspacing}{\spaceskip=\fontdimen2\font plus
\BIBentryALTinterwordstretchfactor\fontdimen3\font minus \fontdimen4\font\relax}
\providecommand{\BIBforeignlanguage}[2]{{%
\expandafter\ifx\csname l@#1\endcsname\relax
\typeout{** WARNING: IEEEtran.bst: No hyphenation pattern has been}%
\typeout{** loaded for the language `#1'. Using the pattern for}%
\typeout{** the default language instead.}%
\else
\language=\csname l@#1\endcsname
\fi
#2}}
\providecommand{\BIBdecl}{\relax}
\BIBdecl

\bibitem{han2024exploring}
B.~Han, Z.~Lv, A.~Jiang, W.~Huang, Z.~Chen, Y.~Deng, J.~Ding, C.~Lu, W.-Q. Zhang, P.~Fan \emph{et~al.}, ``Exploring large scale pre-trained models for robust machine anomalous sound detection,'' in \emph{ICASSP 2024-2024 IEEE International Conference on Acoustics, Speech and Signal Processing (ICASSP)}.\hskip 1em plus 0.5em minus 0.4em\relax IEEE, 2024, pp. 1326--1330.

\bibitem{wilkinghoff2024self}
K.~Wilkinghoff, ``Self-supervised learning for anomalous sound detection,'' in \emph{ICASSP 2024-2024 IEEE International Conference on Acoustics, Speech and Signal Processing (ICASSP)}.\hskip 1em plus 0.5em minus 0.4em\relax IEEE, 2024, pp. 276--280.

\bibitem{zhang2024dual}
Y.~Zhang, J.~Liu, Y.~Tian, H.~Liu, and M.~Li, ``A dual-path framework with frequency-and-time excited network for anomalous sound detection,'' in \emph{ICASSP 2024-2024 IEEE International Conference on Acoustics, Speech and Signal Processing (ICASSP)}.\hskip 1em plus 0.5em minus 0.4em\relax IEEE, 2024, pp. 1266--1270.

\bibitem{hou2023decoupling}
Q.~Hou, A.~Jiang, W.-Q. Zhang, P.~Fan, and J.~Liu, ``Decoupling detectors for scalable anomaly detection in aiot systems with multiple machines,'' in \emph{2023 IEEE Global Communications Conference (GLOBECOM)}.\hskip 1em plus 0.5em minus 0.4em\relax IEEE, 2023, pp. 1--6.

\bibitem{jiang2023unsupervised}
A.~Jiang, W.-Q. Zhang, Y.~Deng, P.~Fan, and J.~Liu, ``Unsupervised anomaly detection and localization of machine audio: A gan-based approach,'' in \emph{ICASSP 2023-2023 IEEE International Conference on Acoustics, Speech and Signal Processing (ICASSP)}.\hskip 1em plus 0.5em minus 0.4em\relax IEEE, 2023, pp. 1--5.

\bibitem{liu2022anomalous}
Y.~Liu, J.~Guan, Q.~Zhu, and W.~Wang, ``Anomalous sound detection using spectral-temporal information fusion,'' in \emph{ICASSP 2022-2022 IEEE International Conference on Acoustics, Speech and Signal Processing (ICASSP)}.\hskip 1em plus 0.5em minus 0.4em\relax IEEE, 2022, pp. 816--820.

\bibitem{wilkinghoff2021sub}
K.~Wilkinghoff, ``Sub-cluster adacos: Learning representations for anomalous sound detection,'' in \emph{2021 International Joint Conference on Neural Networks (IJCNN)}.\hskip 1em plus 0.5em minus 0.4em\relax IEEE, 2021, pp. 1--8.

\bibitem{mcmahan2017communication}
B.~McMahan, E.~Moore, D.~Ramage, S.~Hampson, and B.~A. y~Arcas, ``Communication-efficient learning of deep networks from decentralized data,'' in \emph{Artificial intelligence and statistics}.\hskip 1em plus 0.5em minus 0.4em\relax PMLR, 2017, pp. 1273--1282.

\bibitem{Dohi_arXiv2023_01}
K.~Dohi, K.~Imoto, N.~Harada, D.~Niizumi, Y.~Koizumi, T.~Nishida, H.~Purohit, R.~Tanabe, T.~Endo, and Y.~Kawaguchi, ``Description and discussion on {DCASE} 2023 challenge task 2: First-shot unsupervised anomalous sound detection for machine condition monitoring,'' \emph{In arXiv e-prints: 2305.07828}, 2023.

\bibitem{JieIESEFPT2023}
J.~Jie, ``Anomalous sound detection based on self-supervised learning,'' DCASE2023 Challenge, Tech. Rep., June 2023.

\bibitem{LvHUAKONG2023}
Z.~Lv, B.~Han, Z.~Chen, Y.~Qian, J.~Ding, and J.~Liu, ``Unsupervised anomalous detection based on unsupervised pretrained models,'' DCASE2023 Challenge, Tech. Rep., June 2023.

\bibitem{ramaswamy2000efficient}
S.~Ramaswamy, R.~Rastogi, and K.~Shim, ``Efficient algorithms for mining outliers from large data sets,'' in \emph{Proc. 2000 ACM SIGMOD Int. Conf. Manag. Data}, 2000, pp. 427--438.

\bibitem{dohi2021flow}
K.~Dohi, T.~Endo, H.~Purohit, R.~Tanabe, and Y.~Kawaguchi, ``Flow-based self-supervised density estimation for anomalous sound detection,'' in \emph{ICASSP 2021-2021 Proc. IEEE Int. Conf. Acoust., Speech, Signal Process. (ICASSP)}.\hskip 1em plus 0.5em minus 0.4em\relax IEEE, 2021, pp. 336--340.

\bibitem{li2019abnormal}
S.~Li, Y.~Cheng, Y.~Liu, W.~Wang, and T.~Chen, ``Abnormal client behavior detection in federated learning,'' \emph{arXiv preprint arXiv:1910.09933}, 2019.

\bibitem{nguyen2019diot}
T.~D. Nguyen, S.~Marchal, M.~Miettinen, H.~Fereidooni, N.~Asokan, and A.-R. Sadeghi, ``D{\"i}ot: A federated self-learning anomaly detection system for iot,'' in \emph{2019 IEEE 39th International conference on distributed computing systems (ICDCS)}.\hskip 1em plus 0.5em minus 0.4em\relax IEEE, 2019, pp. 756--767.

\bibitem{park19e_interspeech}
D.~S. Park, W.~Chan, Y.~Zhang, C.-C. Chiu, B.~Zoph, E.~D. Cubuk, and Q.~V. Le, ``{SpecAugment: A Simple Data Augmentation Method for Automatic Speech Recognition},'' in \emph{Proc. Interspeech 2019}, 2019, pp. 2613--2617.

\bibitem{dosovitskiy2021an}
A.~Dosovitskiy, L.~Beyer, A.~Kolesnikov, D.~Weissenborn, X.~Zhai, T.~Unterthiner, M.~Dehghani, M.~Minderer, G.~Heigold, S.~Gelly, J.~Uszkoreit, and N.~Houlsby, ``An image is worth 16x16 words: Transformers for image recognition at scale,'' in \emph{International Conference on Learning Representations}, 2021.

\bibitem{vaswani2017attention}
A.~Vaswani, N.~Shazeer, N.~Parmar, J.~Uszkoreit, L.~Jones, A.~N. Gomez, {\L}.~Kaiser, and I.~Polosukhin, ``Attention is all you need,'' \emph{Advances in neural information processing systems}, vol.~30, 2017.

\bibitem{dawalatabad21_interspeech}
N.~Dawalatabad, M.~Ravanelli, F.~Grondin, J.~Thienpondt, B.~Desplanques, and H.~Na, ``{ECAPA-TDNN Embeddings for Speaker Diarization},'' in \emph{Proc. Interspeech 2021}, 2021, pp. 3560--3564.

\bibitem{pmlr-v202-chen23ag}
S.~Chen, Y.~Wu, C.~Wang, S.~Liu, D.~Tompkins, Z.~Chen, W.~Che, X.~Yu, and F.~Wei, ``{BEAT}s: Audio pre-training with acoustic tokenizers,'' in \emph{Proceedings of the 40th International Conference on Machine Learning}, ser. Proceedings of Machine Learning Research, vol. 202.\hskip 1em plus 0.5em minus 0.4em\relax PMLR, 23--29 Jul 2023, pp. 5178--5193.

\bibitem{deng2019arcface}
J.~Deng, J.~Guo, N.~Xue, and S.~Zafeiriou, ``Arcface: Additive angular margin loss for deep face recognition,'' in \emph{Proceedings of the IEEE/CVF conference on computer vision and pattern recognition}, 2019, pp. 4690--4699.

\bibitem{shi2024sam}
Y.~Shi, P.~Fan, Z.~Zhu, C.~Peng, F.~Wang, and K.~B. Letaief, ``Sam: An efficient approach with selective aggregation of models in federated learning,'' \emph{IEEE Internet of Things Journal}, 2024.

\bibitem{jiang2023thuee}
A.~Jiang, Q.~Hou, J.~Liu, P.~Fan, J.~Ma, C.~Lu, Y.~Zhai, Y.~Deng, and W.-Q. Zhang, ``Thuee system for first-shot unsupervised anomalous sound detection for machine condition monitoring,'' \emph{Proceedings of the IEEE AASP Challenge on Detection and Classification of Acoustic Scenes and Events, Tampere, Finland}, pp. 20--22, 2023.

\bibitem{loshchilov2018decoupled}
I.~Loshchilov and F.~Hutter, ``Decoupled weight decay regularization,'' in \emph{International Conference on Learning Representations}, 2019.

\end{thebibliography}

\end{document}